\documentclass[aps,pra,floatfix,showpacs,noshowkeys,epsfig,graphics,natbib]{revtex4}%
\usepackage{graphicx}
\usepackage{amsmath}
\usepackage{amsfonts}
\usepackage{amssymb}

\newcommand{\newc}{\newcommand}
\newc{\beq}    {\begin{equation}}
\newc{\eeq}    {\end{equation}}

\newc{\beqa}    {\begin{eqnarray}}
\newc{\eeqa}    {\end{eqnarray}}
\newc{\bs}    {\section}
\newc{\no}    {\\ \nonumber}

\newc{\st}    {\stackrel}
\newc{\dsp}{\displaystyle}
\newc{\nn}  {\nonumber}

\begin{document}

\title{Dark energy, inflation and the cosmic coincidence problem}

\author{Jungjai Lee}
\email{jjlee@daejin.ac.kr}
\affiliation{Department of Physics, Daejin University, Pocheon, Gyeonggi 487-711, Korea}

\author{Hyeong-Chan Kim}
\affiliation{Department of Physics, Yonsei University,
Seoul 120-749, Republic of Korea.
}%

\author{Jae-Weon Lee}
\affiliation{School of Computational Sciences,
             Korea Institute for Advanced Study,
             207-43 Cheongnyangni 2-dong, Dongdaemun-gu, Seoul 130-012, Korea}

\date{\today}
\begin{abstract}
  We show that  holographic dark energy could explain why the current dark energy density is
so small, if there was
 an inflation with a sufficient expansion in the early universe.
It is also suggested that  an inflation with the number of e-folds $N\simeq 65$
may  solve the cosmic coincidence problem in this context.
Assuming  the inflation  and the power-law acceleration  phase today we obtain
 approximate formulas for  the event horizon size of the universe  and
dark energy density as  functions of time.
A simple numerical study exploiting the  formula
  well reproduces the observed evolution of dark energy.
This nontrivial match between the theory and the observational data
supports both  inflation and
 holographic dark energy models.
\end{abstract}

\maketitle

The type Ia supernova (SN Ia)
observations~~\cite{riess-1998-116,perlmutter-1999-517}
strongly suggest  that the  current universe is in an  accelerating phase,
which can be explained by  dark
energy ( a generalization of the cosmological constant) having  pressure $p_\Lambda$ and density
$\rho_\Lambda$ such that $\omega_\Lambda\equiv
p_\Lambda/\rho_\Lambda<-1/3$.
 There are various dark energy models rely on exotic materials such as
quintessence~\cite{PhysRevLett.80.1582,PhysRevD.37.3406},
$k$-essence~~\cite{kessence,PhysRevLett.85.4438},
phantom~~\cite{phantom}, and
Chaplygin gas~~\cite{Chaplygin,Bilic:2002vm}.
Being one of the most important unsolved puzzles in modern
physics, the cosmological constant problem consists of three sub-problems;
why the cosmological constant is so small, nonzero, and comparable to
the critical density at the present.

In this paper we show that, in the holographic dark energy model, an inflation with a sufficient expansion
explain why the current dark energy density is so small. We also
suggest that the last problem, the cosmic coincidence problem,
could be  solved,
if there was an inflation with a specific expansion.
Note that, in many other dark energy models, it is not easy
to explain the current ratio  of   dark energy density to  matter energy density,
because usually dark energy density and matter energy density reduce
at different rates~\cite{tracker} for a long cosmological time scale.

It is well known~\cite{DEreview} that a simple combination
of the reduced Planck mass $M_P=m_P/\sqrt{8\pi}$ and the Hubble parameter $H=H_0\sim 10^{-33}~eV$,
gives a value $\rho_\Lambda \simeq M_P^2 H_0^2$ comparable to the observed  dark energy density $\sim
10^{-10} eV^4$ ~\cite{perlmutter-1999-517}.
This interesting coincidence, on one hand, is of the cosmic coincidence problem and,
on the other hand, motivated  holographic dark energy models.
The holographic dark energy models are based on the holographic principle
proposed by 't Hooft and Susskind~\cite{hooft-1993,susskind-1995-36,holography},
 claiming
 that all of the information in a volume  can be described by the physics
  at the boundary
 of the volume. With the base on the principle,
 Cohen et al~\cite{PhysRevLett.82.4971} proposed  a relation between an UV cutoff ($a$) and an IR cutoff ($L$)
by considering that
the total energy in a region of size $L$ can not be larger than
the mass of a black hole of that size. Saturating the bound, one can obtain
 \beq
 \label{holodark}
\rho_\Lambda=\frac{3 d^2}{ L^2 a^2},
 \eeq
where $d$ is a constant.
 Hsu~\cite{hsu} pointed out that for $L=H^{-1}$, the holographic dark energy behaves like
matter  rather than dark energy.
Many attempts~\cite{horvat:087301,myung-2005-610,myung-2005-626,0295-5075-71-5-712,gong:064029,pavon-2005-628}
have been made
  to overcome this IR cutoff problem,  for example,
  by using non-minimal coupling to a scalar field
~\cite{0295-5075-71-5-712,gong:064029} or an interaction between dark energy
and dark matter ~\cite{intDE,pavon-2005-628,Setare:2006wh,Setare:2006sv,Setare:2007at}.
Li~\cite{Li:2004rb,1475-7516-2004-08-013} suggested  that an
ansatz for the
 holographic dark energy density
 \beq
 \label{holodark2}
\rho_\Lambda=\frac{3 d^2 M_P^2}{ R_h^2 },
\eeq
 would  give a correct accelerating universe, where   the future event horizon ($R_h$) is used instead
 of  the Hubble horizon
 as the IR cutoff $L$.

To solve the coincidence problem many attempts have been done
\cite{Ishwaree,Leith:2007bu,RefWorks:5,li-2004-603,Sadjadi:2007ts,Setare:2006wh,RefWorks:5,Kim:2005at}.
An interaction of dark matter~\cite{darkmatter} with  dark energy
was introduced in ~\cite{intDE,Gumjudpai:2005ry,Elizalde:2005ju}.
In ~\cite{li-2004-603} inflation at the GUT scale with the minimal number of e-folds $N\simeq 60$
was suggested as a solution.
In this paper  we suggest a solution similar to the later.
One motivation to study the cosmic coincidence problem in the context of inflationary  cosmology is that
 if there was no inflation, there could be no `now' ($t_0=1.37\times 10^{10}$ years)
 for the `why now' question.
According to astronomical observations and cosmological theory there are at least
two inflationary periods in the history of the universe.
As is well known, the first
inflation at the early universe with $N > 60$ is need to  solve
the  problems of the standard big-bang cosmology.
This inflation  is often assumed to be related to  vacuum energy of a scalar field (inflaton).
The second inflation (re-inflation) is a period of an accelerated  expansion today
 due to dark energy.
(Usually,  the first inflation is related to a phase transition of the inflaton
and  has a different origin
from that of the re-inflation due to dark energy.  In this paper we assume this case. )
Thus, we assume that in the universe there are the inflaton, holographic dark energy, radiation and matter (mostly, cold dark matter).
We also assume that after reheating  inflaton energy  decays to radiation perfectly.
During the first inflation holographic dark energy is diluted exponentially.
In this work we suggest that
if there is  holographic dark energy in the universe, the first inflation with $N\simeq 65$
leads to onset of the second inflation at the time $t_a=O(10^{9})$ years as observed,
and, hence, the inflation solves the cosmic coincidence problem in the context of holographic dark energy.

In this paper we consider the flat ($k=0$)
Friedmann universe which is favored by observations~\cite{wmap} and described by the metric
\beq
ds^2=-dt^2+R^2(t)d\Omega^2,
 \eeq
 where $R(t)$ is the scale factor.
In the holographic dark energy model a typical length scale of the
system with the horizon is given by
 the future event horizon
 \beq
\label{Rh}
R_h\equiv R(t)\int_t^\infty \frac{d R(t')}{H(t')R(t')^2}
= R(t)\int_t^\infty \frac{dt'}{R(t')},
\eeq
which is a key quantity.
It is a subtle task to obtain an explicit form for $R_h(t)$, because
$R_h(t)$ depends on the whole history of the universe after $t$.
To tackle this problem
we divide the history of the universe into two phases;  the inflation (phase 1)
 is followed
by phase 2 which are consecutive
 radiation dominated era (RDE; $R(t)\propto t^{1/2}$) and
 dark energy dominated
power-law accelerating  era (DDE; $R(t)\propto t^{n},~n>1$), respectively.
For simplicity, we  ignore the matter
dominated era (MDE) as often done in an order of magnitude
estimate  in cosmology.
( In appendix, we perform a similar calculation with MDE.
The main results are similar.)

1) inflation phase ( $t_i \le  t < t_f$)\\
The inflation starts at $t=t_i$ and ends at $t_f$.
The scale factor evolves in this phase as follows
 \beq
 R(t)= R_i e^{H_i (t-t_i)},
\eeq
where $R_i$ is the initial scale factor
at $t=t_i$ and  $H_i=M_i^2/(\sqrt{3} M_P)$ is the Hubble parameter with the energy
scale $M_i$ of the inflation.
Hence, the number of e-folds of expansion $N\equiv H_i (t_f-t_i)$.

2) power-law expansion  phase ( $t_f \le t < \infty$)\\
This phase consists of RDE ( $t_f \le t < t_a$)
followed by DDE ( $t_a \le t < \infty$).
The universe starts to accelerate at an inflection point $t=t_a$, i.e., $\ddot{R}(t_a)= 0$.
We assume that the scale factor evolves in this phase as
\beq \label{Rt2}
R(t)= R_i e^{N}\left(\frac{t}{t_f}\right)^{\frac{1}{2}}
    \left(\frac{1+\alpha
\left(\frac{t}{t_f}\right)^{1/2}}{1+\alpha}\right)^{2n},
\eeq
 where $\alpha\simeq(t_f/t_a)^{1/2} $ is a
constant. The scale factor $R(t)$ grows as $t^{\frac{1}{2}}$ during the RDE and as
$t^{n+\frac{1}{2}}$ during the DDE later.
 $R(t)$ of this form gives a smooth transition
from RDE to DDE.
Note that $R(t)$ for each era is well-known
and can be derived from the Friedmann equation depending on the dominant energy source.
The power-law acceleration is a generic feature of DDE if $d>1$.
 (Alternatively, one can divide this phase
into RDE and DDE and choose the scale factor as
$R(t)\propto (t/t_f)^{1/2}$ and $R(t)\propto (t/t_a)^n$ for
RDE and DDE, respectively. This choice gives almost the
same results except for a slightly decreasing $R_h$ as $t\rightarrow t_a$.
Thus, we can use the specific form in Eq. (\ref{Rt2}) without loss of generality.)
Since observational data
favor $d\simeq 1$ \cite{wu2006,zhang-2007} and $\omega_\Lambda$
 close to $-1$, the power index
$n=(1+d)/(2d-2)$
is much larger than $1$. If we choose $d=1.0513$, then $n=20$.
The inflection point $t_a$ is determined by the value at
which the second derivative of $R(t)$ vanishes: \beq
\label{ta:alpha}
 \frac{t_a}{t_f} =
    \frac{1}{\alpha^2}\left(\frac{\sqrt{5n^2-2n}-(n-1)}{
    4n^2-1}\right)^2.
\eeq
From $R(t)$ we obtain $R_h(t)$ using Eq. (\ref{Rh}).
During the inflationary phase (phase 1):
\beqa
 {{I }_1}(t)&\equiv& \int_t^{t_f} \frac{d t'}{R(t')}+\int_{t_f}^{\infty} \frac{d t'}{R(t')}\no
&=& \frac{e^{-H_i(t-t_i) } - e^{-N}}
     {{H_i}\,{R_i}} + C(t_f),
 \eeqa
where $C(t_f)$ is a constant dependent  on $t_f$.
A finite $C(t_f)$ implies a finite  $R_h(t)$ and, hence, the existence of DDE.
This constant should be determined by the initial condition at $t_i$.
Thus the distance to the future horizon $R_h(t)$ during the phase 1 is
\beqa
\label{Rhinflation}
  {R_h}(t)&=&R(t) I_1(t)\no
&=& \frac{1}{H_i}+\left( R_i e^{N} C(t_f)-\frac{1}{H_i}\right) e^{H_i(t-t_f)}.
\eeqa
To determine the value of $C(t_f)$, we use an initial condition $R_h(t_i)$ for $R_h$:
\beq
\label{Rhti}
{R_h}({t_i})=\frac{1}{H_i}+\left( R_i e^{N} C(t_f)-\frac{1}{H_i}\right) e^{-N}.
\eeq
Inserting Eq. (\ref{Rhti}) into Eq. (\ref{Rhinflation}) we can rewrite Eq. (\ref{Rhinflation}) as
\beq
\label{Rhsimple}
R_h(t)= \frac{1}{H_i}\left(1+ A e^{H_i(t-t_i)}\right),
\eeq
where $A$ is a dimensionless constant which depends on the initial condition at $t=t_i$, given by
\beq
A\equiv  H_i R_h(t_i)-1.
\eeq
Therefore, if $H_i R_h (t_i)> 1$, i.e., $A>0$, the event horizon grows exponentially during the inflation.
At the same time $\rho_\Lambda$ decreases exponentially.
This is also noted in Ref. \cite{inflationHDE}, where the correction to the inflation due to holographic
dark energy was investigated.
It is a reasonable assumption that dark energy density at $t_i$ is comparable to other energy densities,
that is, $\rho_\Lambda(t_i)\sim M_P^2/R_h^2\sim H_i^2 M_P^2$, or $R_h(t_i)\sim H_i^{-1}$.
If not, we need  either fine tuning or a special mechanism to make
 initial dark energy density parameter $\Omega_\Lambda$  be much smaller than  1 at the extremely early universe,
 which is implausible.
 This can be also seen from
 the following relation~\cite{1475-7516-2004-08-013};
\beq
H R_h =\frac{d}{\sqrt{\Omega_\Lambda}},
\eeq
which holographic dark energy model should satisfy  all the time.
It is $O(1)$ for $\Omega_\Lambda$ not too much smaller than 1.
Therefore,  $A=O(1)$ is a plausible initial condition.

At $t=t_f$,
\beq \label{Rh:f}
R_h(t_f) =\frac{1}{H_i} (1+ Ae^N) .
\eeq

Now consider the phase 2. Using $R(t)$ in Eq.~(\ref{Rt2}), it is straightforward to obtain the
following relations,
\beqa
\label{I2t}
 {{I }_2}(t)&\equiv& \int_t^{\infty} \frac{d t'}{R(t')}\no
&=&
    \frac{2}{R_i e^{N}} \,
    \frac{(1+\alpha)t_f}{(2n-1)\alpha}
    \left(\frac{1+\alpha\sqrt{\frac{t}{t_f}}}{1+\alpha}
    \right)^{1-2n}.
\eeqa
Therefore, during the phase 2 the event horizon at $t$ is at the distance
\beq \label{Rh:2}
R_h(t) =
    \frac{2t_f}{(2n-1)\alpha}
    \left(\sqrt{\frac{t}{t_f}}+\alpha \frac{t}{t_f}\right).
\eeq
 Now, the horizon distance at $t=t_f$ is
\beq \label{Rhtf}
{R_h}({t_f})=\frac{2(1+\alpha)t_f}{(2n-1)\alpha}.
\eeq
Comparing Eq.~(\ref{Rh:f}) with Eq. ~(\ref{Rhtf}), we have
\beq \label{alpha:N}
 \frac{2(1+\alpha)t_f}{(2n-1)\alpha}=\frac{1}{H_i} (1+Ae^N) .
\eeq
 Therefore, we have
\beq \label{alpha:A}
 \alpha = \left[\frac{n-1/2}{H_i t_f}(1+Ae^N)-1\right]^{-1}
    \simeq
        \frac{H_it_f}{(n-1/2) Ae^N} \ll 1.
\eeq
Inserting
 this into Eq. (\ref{Rh:2}) we obtain
\begin{eqnarray} \label{Rh:t}
R_h(t) =
   \left(\frac{(1+Ae^N)}{H_i
    }-\frac{2t_f}{2n-1} \right)
    \sqrt{\frac{t}{t_f}}+\frac{2t}{2n-1}.
\end{eqnarray}
Now we have   approximate analytical formulas for $R_h(t)$ for the whole history of the universe  since the inflationary era.
 Note that $R_h(t)$ is a monotonically
increasing function of time.
We next consider the behaviors of $R_h(t)$ for $t_f < t\ll t_a$. In this case, we have
\beq
\label{RhRDE}
 R_h(t) \simeq
  t_f\left(\frac{(1+Ae^N)}{H_i
    t_f}-\frac1{n-1/2} \right)
    \sqrt{\frac{t}{t_f}}\simeq
    \frac{Ae^N}{H_i}\sqrt{\frac{t}{t_f}},
 \eeq
which is proportional to $R(t)$.
From  Eq. (\ref{ta:alpha}) the inflection point $t_a$ is given by the initial conditions $A$ and $N$ to be
\beq
\label{tatf}
 \frac{t_a}{t_f} \simeq
 \frac{\left(\sqrt{5n^2-2n}-n+1\right)^2}{4(2n+1)^2}
 \left(\frac{A e^N}{H_i t_f}\right)^2 \simeq 0.095 \left(\frac{A e^N}{N}\right)^2.
\eeq
This relation is interesting and informative.
The ratio of the two time scale $t_a$ and $t_f$ is related to the initial condition.  For $A> 0$
and $N\gg 1$, this ratio explain why dark energy dominates so lately.
Eq. (\ref{tatf}) implies that $t_a$ and, hence, evolution of the universe is more sensitive to $N$ than
 to $A$ or $n$.

From now on all quantities are given in natural units; $m_P=1$.
For $t_f\simeq 10^{7}$ (GUT scale inflation)
 and $N\simeq 66$, a reasonable value for inflation to solve the problems of
the big-bang cosmology, this equation gives observed $t_a\simeq 10^{60}$.
In this way, the holographic dark energy model could solve the cosmic coincidence problem.
Interestingly, Eq. (\ref{tatf})  gives a lower bound for the energy scale of the inflation.
The usual bound $N\st{>}{\sim} 60$ for inflation returns $t_f \st{<}{\sim}10^{12}$ and hence $M_i\st{>}{\sim}
10^{-7} m_P\sim 10^{12}~GeV$.
This can rule out low energy scale inflation models.
On the other hand, an obvious condition $t_f>t_P=1$ returns $N<75$, where $t_P$ is
the Planck time.
To determine the true value of $N$  we need to go beyond the approximation used in this work.

Let us explain more physically how our model could solve the coincidence problem.
Using   Eq. (\ref{Rhsimple}) we obtain
${R_h(t_f)}/{R_h(t_i)}\simeq e^N$,
which means that the event horizon expands
exponentially during the inflation.
At the same time the dark energy density $\rho_\Lambda= 3d^2 M_P^2/R_h^2$ rapidly
decreases (see Fig. \ref{rho});
\beq
\rho_\Lambda (t_f)= \rho_\Lambda (t_i)\left(\frac{R_h(t_i)}{R_h(t_f)}\right)^2
\simeq {M_i^4} e^{-2N},
\eeq
where we used $\rho_\Lambda (t_i)\simeq M_P^2 H_i^2\simeq M_i^4$ and $A\sim O(1)$.
This is the dark energy density just after the inflation.
After the inflation,
 dark energy is sub-dominant, i.e., $\Omega_\Lambda \ll 1$, and behaves like
matter with a constant equation of state~\cite{li-2004-603}
\beq
\label{omega}
\omega_{\Lambda} =-\frac{1}{3} \left(1+\frac{2\sqrt{\Omega_\Lambda} }{d}\right)\simeq -\frac{1}{3}.
\eeq
In this case $\rho_\Lambda\sim R^{-3(1+\omega)}\sim R^{-2}$, while the radiation
energy density,
\beq
\label{rhor}
\rho_r (t) \simeq M_i^4\left(\frac{R(t_f)}{R(t)}\right)^4,
\eeq
decreases more rapidly than the dark energy density.
(From Eq. (\ref{RhRDE}) one can also see that $R_h(t)\propto R(t)\propto t^{1/2}$ during the RDE).
Here we assume an instant reheating after the inflation for simplicity.
Therefore, during the RDE
\beq
\label{rhoLambda}
\rho_\Lambda (t)\simeq  \rho_\Lambda (t_f)\left(\frac{R(t_f)}{R(t)}\right)^2
\simeq
\rho_\Lambda (t_f)\left(\frac{t_f}{t}\right)
= {M^4_i e^{-2N}}\left(\frac{t_f}{t}\right),
\eeq
which should be comparable to $M^4_a$ at $t_a$,
where $M_a\sim 10^{-3} eV$ is the observed energy scale of the universe at
the inflection point $t_a$.
 From the above relation, the required e-folds is
\beq
\label{Napprox}
N\simeq -\frac{1}{2} ln\left[ \left(\frac{t_a}{t_f}\right) \left(\frac{M_a}{M_i}\right)^4\right]
\simeq ln \left(\frac{M_i}{M_a}\right)
\simeq 64.5,
\eeq
which is slightly larger than the minimal $N$ for the inflation to solve the many problems of the standard big bang cosmology.
Here we have used $t_f\sim M_P/M^2_i\sim M_P/(10^{16} GeV)^2$.
This result is comparable with heuristic arguments of Li~\cite{li-2004-603,huang-2004-0408,huang-2005-0503}.
Hence, we see again that an inflation with $N\simeq 65$ could solve the cosmic coincidence problem
in a  self-consistent manner in the holographic dark energy context.

To be more concrete we perform a numerical study using the analytic formulas to fit parameters
for the inflation such as $N$ and $M_i$ onto the observed cosmological parameters such as $\Omega_\Lambda (t_0)$
and $\rho_\Lambda(t_0)$.
Once we know $R_h(t)$ and $R(t)$, it is easy to obtain $\rho_\Lambda(t)$ and
$\rho_r(t)$ by using Eqs. (\ref{rhor}) and (\ref{rhoLambda}) for a given $N$.
We choose reasonable values $M_i=10^{16}~GeV$, $A=1$ and $n=20$.
We will show later that our results are not so sensitive to the value of $A$ or $n$  as long as $n\gg 1$.
From Eqs. (\ref{RhRDE}) and (\ref{tatf}) one can see that $t_a$ and
\beq
\label{Omega}
\Omega_\Lambda(t)\equiv \frac{\rho_\Lambda(t)}{\rho_\Lambda(t)+\rho_r(t)}
\eeq
are sensitive to $N$.

\begin{figure}[htbp]
\includegraphics[width=0.5\textwidth]{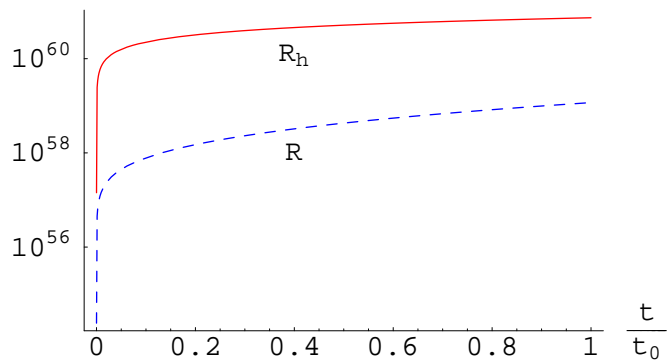}
\caption{(Color online) The size of the event horizon $R_h$ (red thick line)   and
the scale factor $R(t)$ (blue dashed line)   as functions of time $t$
for $N=65.7$, $n=20$, and $M_i=10^{16}~GeV$.
$R_h(t)$ as well as $R(t)$ grows exponentially during the inflation.
All quantities are given in natural units, where $m_P=1$.
\label{Rhfig} }
\end{figure}

\begin{figure}[htbp]
\includegraphics[width=0.5\textwidth]{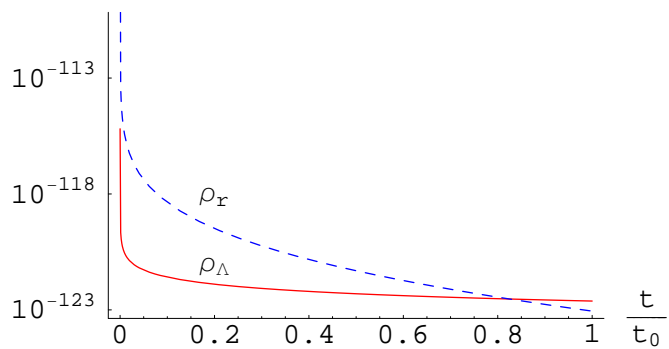}
\caption{(Color online) The dark energy density
$\rho_\Lambda$ (red thick line) and the radiation density $\rho_r$
 (blue dashed line) as a function of time $t$
for the evolution shown in Fig. \ref{Rhfig}.
$\rho_\Lambda(t_0)\simeq 2.4\times 10^{-123}$ is comparable to the observed value.
We ignore the matter dominated era for simplicity.
\label{rho}}
\end{figure}

\begin{figure}[htbp]
\includegraphics[width=0.5\textwidth]{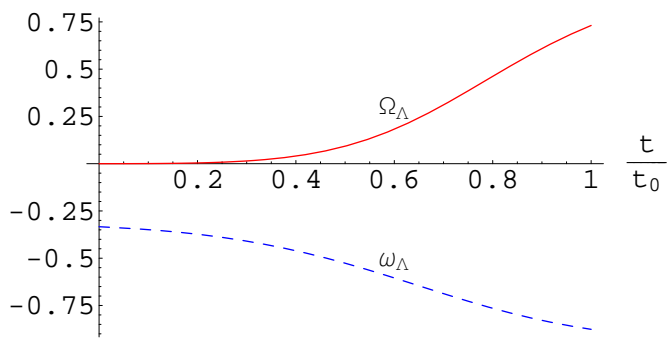}
\caption{(Color online) $\Omega_\Lambda$ (red thick line) from Eq. (\ref{Omega}) and $\omega_\Lambda$
 (blue dashed line) from Eq. (\ref{omega}) of the dark energy as a function of time $t$
for the evolution shown in Fig. \ref{Rhfig}.
$\Omega_\Lambda(t_0)=0.73$ and $\omega_\Lambda(t_0)=-0.876$ are comparable to the observed values.
\label{FigOmega}}
\end{figure}

For $N=65.7$
our theory gives $\Omega_\Lambda(t_0)=0.73$, $\omega_\Lambda=-0.876$, and
$\rho_\Lambda(t_0)\simeq 2.4\times 10^{-123}$, which are comparable with current observations.
 Note that this is not a fine tuning of $N$.
 Since the expansion during the inflation is a history already happened, $N$ is obviously a fixed value.
 Thus, $N$ value is, like other cosmological parameters,
  something which should
   be predicted by a theory and then be verified by observations.
   One can assert fine-tuning only when a required parameter value is unnatural.
    Our model  $predicts$ a  value $~N\simeq 65$ which satisfies all known observational constraints and
is consistent with inflation theory. Thus, the possibility of determining $N$ is not necessary a flaw but a possible merit of our theory.
 Although we can not rule out $N\gg 60$,  interestingly, there is an asserted  upper bound, $N \st{<}{\sim} 65$
from the holographic principle ~\cite{Banks:2003pt,wang:104014,Kaloper:2004gp}  and
from the density perturbation generation
~\cite{PhysRevLett.91.131301,PhysRevD.68.103503}.
If this upper bound is correct, one can say the holographic dark energy can solve the cosmic coincidence problem.

On the other hand, $t_a=0.072~t_0$  is  smaller than the observed value.
This discrepancy can be attributed to   approximations we used such
as an instant reheating after the inflation
and ignoring the matter
dominated era.
If we choose $n=100$ instead of $n=20$, $N=65.715$ gives the same results.
Thus, the results are not sensitive to $n$.
We do not need a fine tuning for $A$ too.
For example, if we choose $A = 10$, then we need $N = 63.39$ to reproduce the observed universe
and $N = 68.01$ for $A = 10^{-1}$. As mentioned above a natural value for this dimensionless
quantity  $A$ without fine tuning is $O(1)$.

Let us recall the inputs and the outputs in our theory.
We have assumed that there are  inflationary era, RDE, and DDE in the observed evolution of our universe
 and used the typical forms of $R(t)$ for these phases.
With reasonable input values for $n,~A$ (our results are not sensitive
to these values)
and $N\simeq 65$,
 we have obtained  output values for current  density
parameter $\Omega_\Lambda (t_0)$ and equation of
state $\omega_\Lambda (t_0)$ for dark energy, which are comparable with observed values.
This could solve the cosmic coincidence problem.
Note that assuming DDE without an appropriate inflation does not automatically explain why $t_a\sim t_0$.
Considering the long time scale involved ($O(10^{10}$ years)) and the difference between time dependency
of the dark energy density and that of the matter density,
    it is  remarkable that with the parameter $N$, and the reasonable assumptions,
our analysis reproduces  the observed universe
with the correct order of magnitude as shown in the figures.
This indicates that the holographic dark energy  models with $d\simeq 1$
are promising candidates for a correct dark energy model.

In summary,  we show that
 an inflation with a sufficient expansion
make the current holographic dark energy density exponentially small.
It is also possible that an inflation of $N\simeq 65$
could solve the cosmic coincidence problem
 without introducing an interaction with dark matter or  modifying
gravity. 
The holographic dark energy models have
 an intrinsic advantage over
non-holographic  models in that
it does not need fine tuning of parameters or  an  $ad~hoc$ mechanism to
cancel the zero-point energy of the vacuum, simply because it
has no $O(M_P^4)$ zero-point vacuum energy from the start. Quantum field theory  over-counts the
 independent physical degrees
of freedom inside the  volume.
Furthermore, as suggested in this paper, the cosmic coincidence problem
could  be also solved  if there was an inflation with $N\simeq 65$.
All these results support not only the inflation theory
but also the holographic dark energy models with $d\simeq 1$.

\appendix

\section{Including MDE}

In this appendix, we investigate the effect of matter dominated era (MDE) on the evolution of
 holographic dark energy.
We assume that there are a period of the inflation followed by
the radiation dominated era (RDE), a slow transition from matter dominated era
to dark energy dominated era (MDE+DDE) of which scale factors are
given by
\begin{eqnarray} \label{Rtnew}
 R(t)=\left\{ \begin{tabular}{ll}
    $\dsp  R_i e^{H_i (t-t_i)}, $&$ t_i \le  t < t_f, $~{\text (Inflation)}\\
    $\dsp R_i e^{N}\left(\frac{t}{t_f}\right)^{\frac{1}{2}},
        $ & $ t_f \leq t < t_{eq}, $~{\text (RDE)} \\
    $\dsp R_i e^{N}\left(\frac{t_{eq}}{t_f}\right)^{\frac{1}{2}}
    \left(\frac{t}{t_{eq}}\right)^{\frac{2}{3}}
    \left(\frac{1+\alpha
\left(\frac{t}{t_{eq}}\right)^{1/3}}{1+\alpha}\right)^{3n},$ &  $
            t_{eq} \leq t ,$~{\text (MDE+DDE)}\\
     \end{tabular}
     \right.
\end{eqnarray}
respectively,
where
$\alpha$ is a constant.
We set the transition from the radiation dominated era (RDE) to the
matter dominated era (MDE) happens at $t=t_{eq}$, the equipartition
time. That is,
$\rho_r(t_{eq})=\rho_m(t_{eq})$, where $\rho_m$ is matter energy density.
 The last phase consists of the matter dominated era (MDE) ( $t_{eq}
\le t < t_a$) followed by a dark energy dominated era (DDE) ( $t_a \le t
< \infty$).
The inflection point $t_a$ is determined by the value in which the
second derivative of $R(t)$ vanishes during the third phase:
\begin{eqnarray} \label{ta:alpha2}
\frac{t_a}{t_{eq}} \simeq \left(\frac{\sqrt{3}-1}{3\alpha n}\right)^3 .
\end{eqnarray}

\begin{figure}[htbp]
\includegraphics[width=0.5\textwidth]{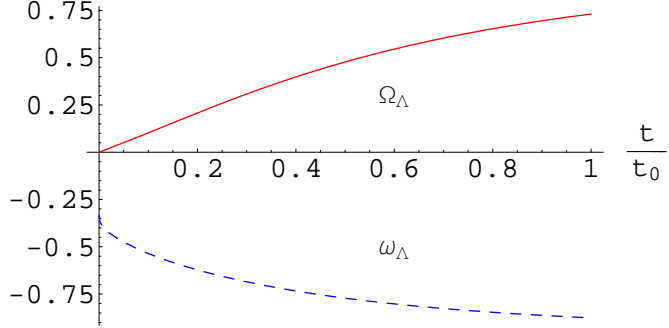}
\caption{(Color online) $\Omega_\Lambda$ (red thick line)  and $\omega_\Lambda$
 (blue dashed line) of the dark energy as a function of time $t$
for the evolution in Eq. (\ref{Rhmatter}), which includes the matter dominated era.
$\Omega_\Lambda(t_0)=0.73$ and $\omega_\Lambda(t_0)=-0.876$.
\label{FigOmega2}}
\end{figure}
\begin{figure}[htbp]
\includegraphics[width=0.5\textwidth]{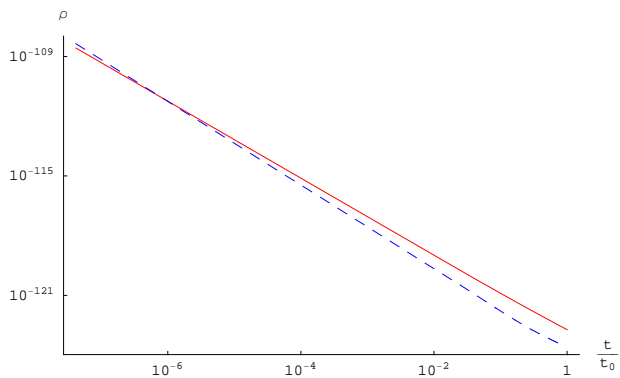}
\caption{(Color online) $3 H^2(t) M_P^2$ (red thick line)  and
the total energy density $\rho_\Lambda(t)+\rho_r(t)+\rho_m(t)$
 (blue dashed line)  as  functions of time $t$
for the evolution in Eq. (\ref{Rhmatter}).
\label{Figcheck}}
\end{figure}

Using the scale factors in Eq. (\ref{Rtnew}), it is easy to derive the event horizon size $R_h$ and dark energy density
$\rho_\Lambda$
by following the procedure in the main text.
After some straightforward calculation we obtain $R_h$ for each phase;
\begin{eqnarray} \label{Rhmatter}
 R_h(t)=\left\{ \begin{tabular}{ll}
    $\dsp \frac{1}{H_i}\left(1+Ae^{H_i(t-t_i)}\right) , $&$ t_i \le  t < t_f, $~{\text (Inflation)}\\
    $\dsp \frac{1+Ae^N+ 2H_i
    t_f}{H_i}\sqrt{\frac{t}{t_f}}-2 t,
        $ & $ t_f \leq t < t_{eq}, $~{\text (RDE)} \\
    $\dsp t_{eq}\left[\left(\frac{1+Ae^N
  +2H_it_f }{H_i\sqrt{t_{eq}t_f}}-2
  -\frac{1}{n-1/3}\right) \left(\frac{t}{t_{eq}}\right)^{2/3}
    +\frac{1}{n-1/3}\frac{t}{t_{eq}}\right],$ &  $
            t_{eq} \leq t ,$~{\text (MDE+DDE)}.\\
     \end{tabular}
     \right.
\end{eqnarray}
During the calculation, from the continuity condition of $R_h$ between RDE and MDE, we have obtained
\beq
 \alpha= \left[\frac{1+Ae^N+ 2H_i t_f
    }{H_i\sqrt{t_{eq}t_f}}
  -2 -\frac{1}{n-1/3}
\right]^{-1}\frac{1}{n-1/3}
\eeq
 and used this $\alpha$ for following calculations.
In generic holographic dark energy models, dark matter is independent of dark energy, and
we need an parameter describing the nature of dark  matter.
We choose the observed equipartition time $t_{eq}\simeq 10^{-7}t_0~(z_{eq}\simeq 3200)$ for the parameter.
Fig. \ref{FigOmega2} shows the results for $N=61.9$.
$N$ becomes smaller compared to the case in Fig. 3 , because the matter energy density decreases slowly
($\rho_m\propto R^{-3}$) than the radiation energy density ($\rho_r\propto R^{-4}$).
The other parameters are the same as those of Fig. 3.
As assumed in the main text, including MDE in our consideration does not significantly change the results.
Compared to the case without MDE (Fig. \ref{FigOmega}),  $\Omega_\Lambda(t)
\equiv \rho_\Lambda(t)/(\rho_\Lambda(t)+\rho_r(t)+\rho_m(t))$ curve is more
flat and $t_a\simeq 5\times 10^9$ years is later. These results are more consistent with   observations, while
 $\rho_\Lambda(t_0)\simeq 2.3\times 10^{-124}$ is slightly smaller than the observed value.
Since our holographic dark energy density changes about $10^{107}$  times in scale from the inflation to the present, this
level of coincidence is interesting, considering the approximations we have used.
 To check the accuracy of our calculation using the Friedmann
equation, we plot the total energy density and $3H^2(t) M_P^2$ in Fig. \ref{Figcheck}.
The graph shows the level of accuracy mentioned above.

 In the case with MDE considered in this appendix, due to the freedom of $t_{eq}$,
 there was no guarantee that  the inflation with $N\sim 65$ could  solve
the cosmic coincidence problem. However, interestingly,  it turns out that even in this case
the required $N$ value is similar to that of the case without MDE. This is due to the fact that the observed initial
dark matter density is much smaller than that of radiation.  Even in the worst case that  MDE started just after
 the  reheating of the inflation and there was no RDE, Eq. (\ref{rhoLambda}) with $R\sim t^{2/3}$
  gives a value $N\simeq \frac{2}{3}
ln \left(\frac{M_i}{M_a}\right)\sim 43$.

\begin{acknowledgments}
Authors are thankful to Changbom Park,
 Yun Soo Myung, Kang Young Lee, and Yeong Gyun Kim for
helpful discussions.
This work was supported by Daejin University Research Grants in 2007.
\end{acknowledgments}


\end{document}